\documentclass[pre,twocolumn,amsmath,amssymb,nofootinbib,floatfix,superscriptaddress]{revtex4}

\usepackage{graphicx,bm,xcolor, bbold}
\usepackage{enumerate}
\usepackage{graphicx,bm}
\usepackage[normalem]{ulem} 

\makeatletter
\def\graphicscale{\twocolumn@sw{0.3}{0.4}}
\def\graphicthreescale{\twocolumn@sw{0.3}{0.4}}

\begin{document}

\title{Out-of-equilibrium scaling of the energy density along the critical
  relaxational flow \\ after a quench of the temperature}

\author{Haralambos Panagopoulos} 
\affiliation{Department of Physics, University of Cyprus,
P.O. Box 20537, 1678 Nicosia, Cyprus}

\author{Ettore Vicari} 
\affiliation{Dipartimento di Fisica dell'Universit\`a di Pisa,
        Largo Pontecorvo 3, I-56127 Pisa, Italy}

\date{\today}

\begin{abstract}
We study the out-of-equilibrium behavior of statistical systems along
critical relaxational flows arising from instantaneous quenches of the
temperature $T$ to the critical point $T_c$, starting from equilibrium
conditions at time $t=0$. In the case of soft quenches, i.e. when the
initial temperature $T$ is assumed sufficiently close to $T_c$ (to
keep the system within the critical regime), the critical modes
develop an out-of-equilibrium finite-size scaling (FSS) behavior in
terms of the rescaled time variable $\Theta=t/L^z$, where $t$ is the
time interval after quenching, $L$ is the size of the system, and $z$
is the dynamic exponent associated with the dynamics. However, the
realization of this picture is less clear when considering the energy
density, whose equilibrium scaling behavior (corresponding to the
starting point of the relaxational flow) is generally dominated by a
temperature-dependent regular background term or mixing with the
identity operator. These issues are investigated by numerical analyses
within the three-dimensional lattice $N$-vector models, for $N=3$ and
$N=4$, which provide examples of critical behaviors with negative
values of the specific-heat critical exponent $\alpha$, implying that
also the critical behavior of the specific heat gets hidden by the
background term.  The results show that, after subtraction of its
asymptotic critical value at $T_c$, the energy density develops an
asymptotic out-of-equilibrium FSS in terms of $\Theta$ as well, whose
scaling function appears singular in the small-$\Theta$ limit.
\end{abstract}

\maketitle


\section{Introduction}
\label{intro}

Various intriguing out-of-equilibrium phenomena are observed in
many-body systems when they are close to a phase transition, where
large-scale modes equilibrate slowly. We mention hysteresis, ageing,
out-of-equilibrium scaling behavior, defect production, etc., which
have been addressed both theoretically and experimentally, at
classical and quantum phase transitions (see, e.g.,
Refs.~\cite{Kibble-80,Binder-87,Bray-94,Zurek-96,CG-05,BDS-06,
  Dziarmaga-10,PSSV-11,Biroli-16,PV-17,RV-21} and references therein).
Understanding the universal features of the out-of-equilibrium scaling
properties at phase transitions is a great challenge within
statistical mechanics and condended matter physics.  They are often
studied considering dynamic protocols based on instantaneous quenches
of the model parameters, see
e.g. Refs.~\cite{CG-05,PEF-12,GS-12,CC-16,PRV-18,RV-20,RV-21}, or
protocols entailing slow changes of the model parameters like those
associated with the so-called Kibble-Zurek problem, see
e.g. Refs.~\cite{Kibble-80,Zurek-96,PSSV-11,CEGS-12,Biroli-16,RV-21,DV-23}.

In this paper we study some aspects of the out-of-equilibrium dynamics
arising from instantaneous changes (quenches) of the temperature to
the critical point of continuous phase transitions.  Analogous
protocols have been considered to investigate some interesting
out-of-equilibrium phenomena related to the critical relaxation from
high-temperature conditions, such as ageing, short-time critical
dynamics, etc., see, e.g.,
Refs.~\cite{CG-05,Suzuki-76,Suzuki-77,JSS-89,OJ-95,LSZ-95,HPV-07,OI-07,
  Collura-10}.  Here, we focus on soft quenches, for which the
temperature variation associated with the quench is sufficiently small
to maintain the system close to criticality along the whole
post-quench out-of-equilibrium dynamics, giving rise to a critical
relaxational flow. We address the behavior of the energy density along
the critical relaxational flow at thermal transitions.

In statistical systems with short-ranged interactions and of
finite-size $L$, the critical relaxational flow of the critical
large-distance modes is expected to be described by an extension of
the equilibrium finite-size scaling (FSS) behaviors, by simply adding
a further dependence of the scaling functions on the time $t$ after
quench, through an appropriate time scaling variable $t/L^z$, where
$z$ is the universal dynamic exponent associated with the type of
dynamics considered~\cite{Ma-book,HH-77,FM-06}.  In particular, this
is expected to work for observables that have a well-defined
asymptotic equilibrium FSS behavior, such as the two-point function of
the order-parameter field.  See e.g.  Refs.~\cite{PRV-18,RV-21} for an
analysis of analogous behaviors at quantum transitions.

However, the situation becomes less clear in the case of thermodynamic
quantities whose equilibrium behavior at the transition is dominated
by a regular background term, which can be also interpreted as a
mixing with the identity operator in statistical field theory, see
e.g. Refs.~\cite{PV-02,ID-book} and references therein, such as the
energy density at classical transitions driven by temperature
variations. In these cases the singular scaling behavior is only
subleading at criticality, thus hidden by the dominant regular
contributions. For example, we mention the cases of the superfluid
transition in $^4$He and the Curie transition in isotropic
ferromagnets~\cite{Ma-book,PV-02}, where also the specific heat
(temperature derivative of the energy density) is dominated by the
regular term, due to the fact that their critical specific-heat
exponent is negative. Therefore, the energy density and the specific
heat do not generally show an asymptotic equilibrium scaling behavior,
unless quite complicated subtractions of the leading regular terms are
performed. This may be related to the fact that they are essentially
dominated by short-ranged fluctuations, which are weakly coupled to
the long-range critical modes. Therefore the study of their
out-of-equilibrium behavior along critical relaxational flows may
provide interesting information on the interplay between the dynamics
of short and long-ranged critical modes at continuous phase
transitions.

We address this issue by analyzing the out-of-equilibrium behavior of
the energy density arising from instantaneous quenches of the
temperature in statistical systems close to a finite-temperature
criticality. As paradigmatic models we consider the three-dimensional
(3D) lattice $N$-vector models, and present numerical analyses for
$N=3$ and $N=4$, which provide examples of critical behaviors with
negative values of the specific-heat critical exponent $\alpha$.  We
study the out-of-equilibrium behavior along critical relaxational
flows arising from instantaneous quench of the temperature $T$ at a
time $t=0$, from $T>T_c$ to the critical point $T_c$, starting from
equilibrium conditions. Our numerical analyses are performed within
dynamic FSS frameworks. They show that the energy density also
develops a peculiar out-of-equilibrium FSS, after subtraction of its
asymptotic critical value at $T_c$. However, unlike other observables
closely related to the critical modes, such as the two-point function
of the order-parameter field, the out-of-equilibrium FSS of the energy
density presents peculiar singularities. Indeed the corresponding
out-of-equilibrium scaling functions turn out to be singular in the
small rescaled time limit after quenching.

It is worth mentioning that the critical relaxational flow at
finite-temperature transitions is similar to the so-called gradient
flow that is often exploited in four-dimensional lattice quantum
chromodynamics (QCD) that is the theory of strong
interactions~\cite{Creutz-book,MM-book}, to define a running coupling
from the lattice energy density, see
e.g. Refs.~\cite{Luscher-10,LW-11,HN-16}.  Indeed, since the continuum
limit of the lattice QCD is realized in the zero-coupling
(zero-temperature) limit where the lattice length scale diverges
exponentially, the critical (fixed-point) relaxational flow becomes a
simple deterministic gradient flow at vanishing bare gauge coupling
(corresponding to zero temperature), which is equivalent to a Langevin
equation without stochastic term~\cite{Ma-book}. In this case,
perturbative arguments suggest that the energy density along the
gradient flow develops an out-of-equilibrium scaling behavior leading
to a scaling running coupling in the continuum limit. Analogous
behaviors have been put forward for the two-dimensional $N$-vector
models with $N\ge 3$~\cite{MS-15,MSS-15}, whose critical behavior
occurs in the zero-temperature limit like lattice QCD.  Our study
within critical relaxational flows at finite-temperature transitions
may help to clarify some of the basic assumptions of the applications
of the gradient flow in lattice QCD, and in particular the possibility
of extracting a well defined running coupling constant from the energy
density under the gradient flow, beyond perturbation theory.

The paper is organized as follows.  In Sec.~\ref{modprot} we introduce
the lattice $N$-vector models that we consider in this study,
mentioning some of their critical features, and outline the dynamic
protocol to study the critical relaxational flows.  In
Sec.~\ref{dynsca} we address the out-of-equilibrium scaling behavior
associated with a soft quench of the temperature to the critical
point, within a FSS framework that extends the renormalization-group
(RG) theory at equilibrium, and we discuss the problematic case of the
energy density. In Sec.~\ref{numres} we present numerical FSS analyses
of Monte Carlo (MC) simulations for the $N=3$ and $N=4$ vector models,
paying particular attention to the behavior of the energy density,
which shows a nontrivial out-of-equilibrium FSS. Finally, in
Sec.~\ref{conclu} we summarize and draw our conclusions.

\section{Models and dynamic protocols}
\label{modprot}

\subsection{The lattice $N$-vector models}
\label{onmod}

As paradigmatic statistical models undergoing continuous transitions,
we consider 3D $N$-vector models defined on cubic
lattices with real-valued $N$-component site variables ${\bm s}_{\bm
  x}$ of unit length (${\bm s}_{\bm x}\cdot {\bm s}_{\bm x}=1$), whose
Hamiltonian reads
\begin{eqnarray}
  H = - J N \sum_{{\bm x},\mu} {\bm s}_{\bm x} \cdot {\bm s}_{{\bm
      x}+\hat{\mu}}
+ \sum_{\bm x} {\bm h}\cdot {\bm s}_{\bm x}.
\label{onmode}
\end{eqnarray}
The equilibrium thermodynamic properties are determined by the
partition function
\begin{equation}
Z = \sum_{\{{\bm s}_{\bm x}\}} e^{-\beta H},\qquad \beta=1/T.
\label{partfunc}
\end{equation}
We set $J=1$ in the following, which implies that the energy scales
are quantified in units of $J$.  For simplicity, we assume systems
with periodic boundary conditions, thus preserving translational
invariance.

The continuous transitions of 3D $N$-vector models are controlled by
two relevant parameters: the inverse-temperature deviation
\begin{equation}
  r \equiv \beta_c-\beta,
  \label{rdef}
\end{equation}
where $\beta_c$ is the transition point, and the homogeneous external
field ${\bm h}\equiv h \hat{n}$ where $\hat{n}$ is a generic direction
of the $O(N)$ symmetry group. The phase transition separates a
paramagnetic phase with $r > 0$ from a ferromagnetic phase with $r <
0$.  When approaching the critical point, the length scale $\xi$ of
the critical modes diverges as $\xi \sim |r|^{-1/y_r}$ for $h=0$, and
as $\xi \sim |h|^{-1/y_h}$ for $r=0$, where $y_r$ and $y_h$ are the RG
dimensions of the parameters $r$ and $h$, respectively (see
e.g. Refs.~\cite{ZJ-book,PV-02}).  The universal critical exponents,
$y_r$ and $y_h$, appearing in these power laws, can be related to the
correlation-length exponent $\nu\equiv 1/y_r$, and to the exponent
$\eta$ associated with the power-law decay of the two-point function
$G({\bm x})$ of the order-parameter field at the critical point,
i.e. $G({\bm x})\approx 1/|{\bm x}|^{d-2+\eta}$, related to $y_h$ by
$y_h=(d+2-\eta)/2$.  Other exponents describe the critical power laws
of other observables at equilibrium, but they can be related to $\nu$
and $\eta$ by scaling and hyperscaling relations.

Several accurate results are known for the critical behaviors of
$N$-vector models.  We mention that the 3D Ising ($N=1$) model
develops a continuous transition at $\beta_c =
0.221654626(5)$~\cite{FXL-18}, whose critical exponents are known very
accurately, see, e.g.,
Refs.~\cite{PV-02,GZ-98,CPRV-02,Hasenbusch-10,KPSV-16,KP-17,FXL-18};
in particular we report the estimates~\cite{KPSV-16} $\nu=0.629971(4)$
and $\eta=0.036298(2)$, and $\omega= 0.8297(2)$ for the leading
scaling-correction exponent, see Sec.~\ref{equfss}.  The Heisenberg
($N=3$) model undergoes a continuous transition ar
$\beta_c=0.2310010(7)$~\cite{DBN-05,BFMM-96}.  Accurate estimates of
the critical exponents and other universal features can be found in
Refs.~\cite{Hasenbusch-20-o3,Chester-etal-20-o3,HV-11,CHPRV-02}.  In
particular, we mention the accurate estimates: $\nu = 0.71164(10)$,
$\eta=0.03784(5)$, and the leading scaling-correction exponent
$\omega=0.759(2)$ obtained in Ref.~\cite{Hasenbusch-20-o3}.  The $N=4$
model is known to undergo a continuous transition
at~\cite{Hasenbusch-22} $\beta_c=0.23396363(6)$, with critical
exponents $\nu=0.74817(20)$, $\eta=0.03624(8)$, and $\omega =
0.755(5)$.

The dynamics of many-body systems at continuous transitions is
generally affected by critical slowing down.  Indeed the time scale of
the critical modes tend to diverge as well, as $\tau\sim \xi^z$ where
$z$ is an independent dynamic exponent, whose value also depends on
the type of dynamics~\cite{HH-77,FM-06}. The simplest example is
provided by the purely relaxational dynamics associated with a
stochastic Langevin equation where only dissipative couplings are
present, with no conservation laws, usually called dynamic model A.
Due to the critical slowing down, the thermalization process tends to
slow down when approaching the critical point.

To estimate the purely relaxational dynamic exponent $z$ of the $N=3$
and $N=4$ vector models, we use the field theoretical
$\varepsilon$-expansion result~\cite{FM-06,AV-84,HH-77,HHM-72,PV-16}
\begin{equation}
  z = 2 + c\,\eta,\qquad c = 0.726 - 0.137\varepsilon
  +O(\varepsilon^2),
  \label{zmodela}
  \end{equation}
where the constant $c$ has been estimated by three-loop computations
within the $\varepsilon=4-d$ expansion framework, and it turns out to
be independent of $N$ to the known $O(\varepsilon)$ order. Therefore,
using the above-reported estimates of the critical exponent $\eta$,
and setting $\varepsilon=1$, we estimate $z=2.022(5)$ and $z=2.021(5)$
respectively for the $N=3$ and $N=4$ vector models [where, as estimate
  of the uncertainty, we take the contribution coming from the
  $O(\varepsilon)$ term of $c$]. An analogous procedure would give
$z=2.021(5)$ for $N=1$, in agreement with the more accurate estimate
$z=2.0245(15)$ for the 3D Ising universality class, obtained by
numerical analyses in Ref.~\cite{Hasenbusch-20}.

In numerical analyses of the equilibrium properties, and in particular
in those based on FSS techniques, one considers a number of
observables, whose asymptotic FSS behavior is described by RG scaling
equations, see next section.  The energy density $E$ and specific heat
$C_V$ are defined as
\begin{eqnarray}
  && E =- {1\over L^d} {\partial \ln Z\over \partial \beta}
  = {1\over
    L^d} \langle H \rangle ,\label{enedef}\\ && C_V
 = {1\over L^d} {\partial^2 \ln Z\over \partial \beta^2}
  = {1\over L^d} \left(
  \langle H^2\rangle - \langle H^2\rangle\right).
  \label{cvdef}
\end{eqnarray}
We also consider the two-point spin function
\begin{equation}
G({\bm x}-{\bm
  y})=\langle {\bm s}_{\bm x} \cdot {\bm s}_{\bm y} \rangle,
\label{twopfun}
\end{equation}
from which one can extract the susceptibility $\chi$ and second-moment
correlation length $\xi$, defined as
\begin{eqnarray}
  \chi \equiv  \widetilde{G}({\bm 0}),\quad
  \xi^2 \equiv {1\over 4 \sin^2 (\pi/L)}
     {\widetilde{G}({\bm 0})
       - \widetilde{G}({\bm p}_m)\over 
\widetilde{G}({\bm p}_m)},
\label{xidefpb}
\end{eqnarray}
where $\widetilde{G}({\bm p})=\sum_{{\bm x}} e^{i{\bm p}\cdot {\bm x}}
G({\bm x})$ and ${\bm p}_m=(2\pi/L,0,...)$ is the smallest lattice
momentum taking the minimal value $2\pi/L$ along one direction only.
In FSS studies of critical behaviors, one may also consider RG
invariant quantities whose scaling behavior does not depend on any
nonuniversal normalization, such as the ratio
\begin{equation}
 R \equiv \xi/L, 
 \label{defS}
\end{equation}
and the Binder parameter defined as
\begin{equation}
  U \equiv {\langle m_{2}^2\rangle\over \langle m_{2} \rangle^2}, \qquad
  m_{2} \equiv {1\over L^d} \sum_{{\bm x},{\bm y}} {\bm s}_{\bm x} \cdot
  {\bm s}_{\bm y}.
  \label{defU}
\end{equation}

\subsection{Protocol to study the critical relaxational flow}
\label{protflow}

We want to study the dynamic scaling behavior under an
out-of-equilibrium critical relaxational flow. For simplicity, we
assume that the external field ${\bm h}$ vanishes (the results of our
analyses can be straightforwardly extended to allow for a nonzero
${\bm h}$). We study the out-of-equilibrium relaxational flow arising
from an instantaneous quench of the inverse temperature, from
$\beta<\beta_c$ to $\beta_c$. In practice, we consider the following
protocol:

(i) We start from a Gibbs ensemble of equilibrium configurations,
equilibrated at an inverse temperature $\beta<\beta_c$, which can be
easily obtained by standard MC techniques at equilibrium.

(ii) These configurations are the starting point for an
out-equilibrium critical relaxational flow at the critical point
$\beta_c$, starting at $t=0$.  This is achieved by making the system
evolve using a purely critical relaxational dynamics at fixed
$\beta_c$, such as that arising from a purely relaxational Langevin
equation or upgradings based on Metropolis and heat-bath algorithms in
lattice systems, without any conservation law (model A of the
dynamics~\cite{HH-77}). 

The out-of-equilibrium dynamics arising from this quench protocol can
be monitored using the quantities analogous to those introduced in
Sec.~\ref{onmod}, by computing them at fixed time $t$, such as the
instantaneous energy density $E$ defined from the expectation value of
$H$, the susceptibility $\chi$, the correlation length $\xi$, ratio
$R=\xi/L$ and Binder parameter $U$.  They are obtained by averaging
over the ensemble of configurations at the time $t$ of the critical
relaxational flow.

The above dynamic protocol allows us to study the critical
relaxational flow in out-of-equilibrium conditions, arising from an
initial sudden quench from $\beta<\beta_c$ to $\beta_c$, i.e., from
$r=\beta_c-\beta>0$ to $r=0$.  Analogous protocols have been often
considered to investigate ageing phenomena, see
e.g. Ref.~\cite{CG-05}. In this paper we consider soft quenches,
i.e. when starting from initial conditions close to the critical
point, i.e. for small values of the inverse-temperature deviation $r$,
so that the system stays always within the critical regime during the
relaxational flow.  Of course, for finite-size systems the equilibrium
is eventually recovered for large times, which tend to be larger and
larger with increasing $L$.

As previously mentioned, the above protocol, monitoring the behavior
of the system along a purely relaxational flow, is the analog of the
so-called gradient flow widely used in lattice QCD to define scaling
observables, and in particular to define a scaling coupling constant
from the energy density~\cite{Luscher-10,LW-11,MS-15,MSS-15,HN-16}.
In lattice QCD one performs a quench from a finite bare coupling $g_0$
to the critical value $g_0=0$, thus corresponding to a simple gradient
flow without requiring stochastic terms, unlike the relaxational flow
at the finite critical temperature.  As we shall see,
finite-temperature transitions show a more complex scenario, related
to the fact that the critical relaxational flow is driven by a
stochastic dynamics tending to thermalize the system at a critical
finite temperature, asymptotically in the large-time limit.

\section{Out-of-equilibrium scaling along the critical relaxational
flow}
\label{dynsca}

In this section we discuss the out-of-equilibrium scaling behavior
associated with the protocol outlined in Sec.~\ref{protflow}, within
an out-of-equilibrium FSS framework. For simplicity we assume
finite-size systems of size $L$ without boundaries, thus preserving
translational invariance, such as those subject to periodic boundary
conditions.

We recall that FSS describes the critical behavior when the
correlation length $\xi$ of the critical modes becomes comparable to
the size $L$ of the system.  The FSS limit can be defined as the
large-$L$ limit, keeping the ratio $\xi/L$ fixed.  This regime
presents universal features, shared by all systems whose transition
belongs to the same universality class, see
e.g. Refs.~\cite{FB-72,Barber-83,Privman-90,Cardy-editor,PV-02,CPV-14,RV-21}.
FSS can be also extended to dynamic phenomena, by also performing an
appropriate rescaling of the time, i.e. supplementing the equilibrium
FSS with the further condition of keeping the ratio $t/L^z$ fixed,
where $z$ is an universal dynamic exponent, see, e.g.,
Refs.~\cite{Suzuki-76,Suzuki-77,JSS-89,OJ-95,LSZ-95,PV-02,
  CG-05,HPV-07,OI-07,Collura-10,PRV-18,RV-21}.

\subsection{Equilibrium finite-size scaling}
\label{equfss}

Before discussing the out-of-equilibrium scaling flow associated with
the protocol outlined in Sec.~\ref{protflow}, we briefly recall the
main features of the equilibrium FSS holding in particular for the
initial equilibrium state before quenching, and in the large-time
limit after the quench (at finite volume).  The free-energy density
can be written as~\cite{WK-74,Fisher-74,Wegner-76,PV-02,RV-21}
\begin{equation}
  F(r,h,L) = F_{\rm reg}(r,h^2) + F_{\rm sing}\big(L,u_r,u_h, \{v_i\} \big)\,,
  \label{Gsing-RG-1fss}
\end{equation}
where $r\equiv \beta_c-\beta$, and $F_{\rm reg}(r,h^2)$ is a
nonuniversal regular function (we neglect its weak dependence on $L$,
which is conjectured to be exponentially suppressed in the large-$L$
limit~\cite{Privman-90,SS-00,PV-02,CPV-14}), and $F_{\rm sing}$ bears
the nonanalyticity of the critical behavior.  The scaling fields $u_r$
and $u_h$ are analytic functions associated with the parameters $r$
and $h$, so that $u_r = r + c_r r^2 + O(r^3,h^2r)$ and $u_h = h + c_h
r h + O(h^3,r^2 h)$, where $c_r$ and $c_h$ are nonuniversal constants.
Besides the relevant scaling fields $u_r$ and $u_h$, there is also an
infinite number of irrelevant scaling fields $\{v_i\}$ (that are also
analytic functions of the system parameters) with RG dimensions
$y_i<0$, giving rise to power-law $O(L^{y_i})$ scaling
corrections to the asymptotic critical behavior.  Using standard
notation, and assuming that they are ordered so that $|y_1| \le |y_2|
\le \ldots$, we define the leading scaling-correction exponent as
$\omega = -y_1$.  The singular part of the equilibrium free energy
show the asymptotic FSS behavior
\begin{eqnarray}
  &&F_{\rm sing} \approx 
  L^{-d} \Bigl[ {\cal F}(\Upsilon, \Gamma) + O(L^{-\omega})\Bigr],
  \label{leadINGFSS}\\
&& \Upsilon = L^{y_r} r, \qquad y_r=1/\nu,\qquad \Gamma = L^{y_h} h.
\nonumber
\end{eqnarray}
Analogous scaling behaviors for the energy density $E$ and specific
heat $C_V$ can be straightforwardly obtained by taking appropriate
derivatives of the free-energy density, cf. Eqs.~(\ref{enedef}) and
(\ref{cvdef}).

Assuming that the external symmetry-breaking field vanishes,
i.e. $h=0$, the equilibrium energy density $E_e\sim \partial
F/\partial r$ is expected to behave as
\begin{eqnarray}
E_e(r,L) \approx E_{\rm reg}(r) + L^{-(d-y_r)} {\cal E}_e(\Upsilon).
  \label{leadINGFSSene}
\end{eqnarray}
The scaling function ${\cal E}_e(Y)$ is expected to be universal apart
from a multiplicative constant and trivial normalizations of the
scaling argument $Y$.  Note, however, that the scaling term turns out
to be subleading with respect to the regular one, due to the fact that
$d-y_r>0$ at continuous transitions ($y_r<d$ otherwise the singularity
becomes inconsistent with a continuous transition). Therefore, the
behavior of the energy density is dominated by the analytical
background, which can be identified as a mixing with the identity
operator.  The specific heat is also dominated by the analytical
background when the corresponding critical exponent $\alpha=2 - d\nu$
is negative~\cite{PV-02}.

To investigate the out-of-equilibrium behavior of the energy density
under the critical relaxational flow associated with the protocol
outlined in Sec.~\ref{protflow}, and in particular its approach to the
equilibrium critical value at $r=0$, we focus on a subtracted energy
density
\begin{eqnarray}
&&  E_{se}(r,L) \equiv E_e(r,L) - E_c,\label{diffe}\\
&&E_c \equiv E_e(r=0,L\to\infty) = E_{\rm reg}(0).
\label{ecdef}
\end{eqnarray}
Therefore $E_{se}(r=0,L\to\infty) = 0$ by definition.  The scaling
behavior of the subtracted energy $E_{se}$ is trivially obtained from
Eq.~(\ref{leadINGFSSene}),
\begin{eqnarray}
  &&E_{se}(r,L) \approx \Delta E_{\rm reg}(r) + L^{-(d-y_r)}
  {\cal
    E}_e(\Upsilon),
  \label{leadINGFSSenesub}\\
&&\Delta E_{\rm reg}(r)\equiv E_{\rm reg}(r) - E_{\rm reg}(0) = b_1 \,r +
  O(r^2),\qquad
 \label{regeexp}
\end{eqnarray}
for generic systems ($b_1$ is a nonuniversal constant).  Note that in
the FSS limit, where $r\sim L^{-y_r}$, the regular term of the
subtracted energy density $E_{se}(r,L)$ is subleading if
\begin{eqnarray}
  2 y_r - d = \frac{2-d\nu}{\nu} = \frac{\alpha}{\nu} > 0,
\label{relcond}
\end{eqnarray}
where $\alpha$ is the specific-heat exponent.  Therefore, for the 3D
Ising models it is subleading, giving only rise to slowly decaying
$O(L^{-\alpha/\nu})$ scaling corrections.  When $\alpha<0$, such as
the case of the $N$-vector models for $N\ge 2$, the regular term is
still dominant, affecting the asymptotic FSS limit which is expressed
as $O(L^{-\alpha/\nu})$. For example, $\alpha/\nu = -0.1896(4)$ for
$N=3$ and $\alpha/\nu=-0.3268(7)$ for $N=4$.

The analytical background does not affect the leading FSS of the other
observables introduced in Sec.~\ref{onmod}.  In particular, the
equilibrium two-point spin correlation $G({\bm x})$ for $h=0$ behaves
as
\begin{equation}
  G_e({\bm x},r,L) \approx L^{-2y_\phi} \,
  {\cal G}_e({\bm X},\Upsilon), \qquad {\bm
    X}={\bm x}/L,
  \label{gxsca}
\end{equation}
where $y_\phi=d-y_h=(d-2+\eta)/2$ is the RG dimension of the
order-parameter field, and ${\cal G}_e({\bm X},\Upsilon)$ is a
corresponding scaling function. Dimensionless RG invariant quantities,
such as $R_e\equiv \xi/L$ and the equilibrium Binder parameter $U_e$,
defined as in Eq.~(\ref{defS}) and (\ref{defU}), are particularly
useful quantities in FSS analyses.  Indeed, their equilibrium scaling
behaviors does not entail power-law prefactors, i.e.
\begin{equation}
  R_e(r,L) \approx {\cal R}_e(\Upsilon),\qquad U_e(r,L) \approx {\cal
    U}_e(\Upsilon),
  \label{rusca}
\end{equation}
where the scaling functions ${\cal R}_e$ and ${\cal U}_e$ are
universal apart from a normalization of the argument $\Upsilon$
(however, we recall that they depend on the type of boundary
conditions and aspect ratio of the finite-size lattice). Scaling
corrections are generally $O(L^{-\omega})$ where $\omega$ is the
leading scaling-correction exponent.

\subsection{Out-of-equilibrium scaling}
\label{outsca}

The dynamics of many-body systems at continuous transitions is
generally affected by critical slowing down. In the FSS framework this
is related to the power-law increase of the time scale $\tau$ of the
equilibrium and out-of-equilibrium evolutions of the critical
long-range modes, as $\tau\sim L^z$ where $z$ is a universal dynamic
exponent associated with the dynamics considered.  Critical dynamic
phenomena crucially depend on the type of dynamics that drives the
many-body system. We consider a purely relaxational dynamics (also
known as model A of critical dynamics~\cite{HH-77,Ma-book,FM-06}),
which can be realized by stochastic Langevin equations, or just
Metropolis or heat-bath updatings in MC simulations, see
e.g. Refs.~\cite{Creutz-book,MM-book,ID-book}.  Within RG frameworks,
the time dependence of correlation functions at different times can be
generally described by adding a further dependence on the scaling
variable $\xi^{-z} t$, or $L^{-z} t$ within FSS.

Concerning the post-quench out-of-equilibrium critical relaxational
flow arising from the protocol outlined in Sec.~\ref{protflow}, the
observables related to the critical modes are expected to develop an
out-of-equilibrium scaling behavior, whose description is achieved by
adding a further dependence on the time scaling variable
\begin{equation}
\Theta = L^{-z} t \sim t/\tau,
\label{thetact}
\end{equation}
similarly to the scaling ansatz for the equilibrium critical dynamics,
see also
Refs.~\cite{CG-05,Suzuki-76,Suzuki-77,JSS-89,OJ-95,LSZ-95,PV-02,
  HPV-07,OI-07,Collura-10}.  Analogous scaling arguments have been
conjectured to describe the out-of-equilibrium behaviors after a
quench at quantum transitions~\cite{PRV-18,RV-21}.

This idea can be straightforwardly applied to the fixed-time two-point
function $G(t,{\bm x},r,L)$ obtained by averages over ensembles at
fixed $t$ along the relaxational flow starting from the equilibrium
configurations at an initial $r\equiv \beta_c-\beta>0$.  According to
the above out-of-equilibrium scaling hypothesis, its
out-of-equilibrium FSS behavior is simply obtained by adding a further
dependence on $\Theta$ to the scaling function ${\cal G}$ entering the
equilibrium FSS reported in Eq.~(\ref{gxsca}). Therefore, we expect
that
\begin{equation}
  G(t,{\bm x},r,L) \approx L^{-2y_\phi} \,{\cal G}({\bm
    X},\Upsilon,\Theta), 
  \label{gxscaout}
\end{equation}
asymptotically in the out-of-equilibrium FSS limit, i.e. the large-$L$
and large-$t$ limits keeping the scaling variables ${\bm X}$,
$\Upsilon$ and $\Theta$ fixed. Analogous behaviors are expected for
the RG invariant quantities $R(t,r,L)$ and $U(t,r,L)$, defined as in
Eqs.~(\ref{defS}) and (\ref{defU}) and measured along the relaxational
flow:
\begin{eqnarray}
  R(t,r,L) \approx {\cal R}(\Upsilon,\Theta), \quad
  U(t,r,L) \approx {\cal U}(\Upsilon,\Theta).
  \label{gxscadyn}
\end{eqnarray}
Since both $R$ and $U$ show already a scaling behavior at $t=0$,
cf. Eq.~(\ref{gxsca}), we may expect a nonsingular $\Theta\to 0$
limit, i.e.
\begin{eqnarray}
{\cal R}(\Upsilon,\Theta\to 0)={\cal R}_e(\Upsilon),\quad {\cal
  U}(\Upsilon,\Theta\to 0)={\cal U}_e(\Upsilon),
  \label{gxscadyneq}
\end{eqnarray}
where ${\cal R}_e$ and ${\cal U}_e$ are the corresponding
equilibrium scaling functions.  Moreover, due to the large-time
thermalization at fixed $L$, we expect that
\begin{eqnarray}
{\cal
  R}(\Upsilon,\Theta\to\infty)\to {\cal R}_e(0),\quad {\cal
  U}(\Upsilon,\Theta\to\infty)\to {\cal U}_e(0).\quad
\label{inflimRU}
\end{eqnarray}
Note that the dependence on $\Upsilon$ is essentially related to the
initial condition of the quench protocol, while no scaling variable is
associated with the post-quench value of $r$ because it is always set
to zero (otherwise a further scaling variable should have been added,
such as $\Upsilon_f\equiv r_f L^{y_r}$ where $r_f$ is the post-quench
value of $r$). The asymptotic out-of-equilibrium FSS is expected to be
approached with power-law scaling corrections, like at equilibrium.
Actually, one may guess that they get generally suppressed as
$L^{-\omega}$, like the leading scaling corrections at equilibrium.

The scaling functions ${\cal R}$ and ${\cal U}$ are expected to be
universal, apart from trivial normalizations of the scaling variables
$\Theta$ and $\Upsilon$.  This implies that, for two different systems
belonging to the same universality class, with the same boundary
conditions, and driven by analogous dynamic types, the corresponding
scaling functions must be related as
\begin{eqnarray}
 {\cal R}(\Upsilon,\Theta)^{(1)} &=& {\cal R}(c_y\Upsilon,c_\theta\Theta)^{(2)},
\label{rescaling}\\
      {\cal U}(\Upsilon,\Theta)^{(1)} &=& {\cal
        U}(c_y\Upsilon,c_\theta\Theta)^{(2)}, \nonumber
\end{eqnarray}
where ${\cal R}(\Upsilon,\Theta)^{(a)}$ and ${\cal
  U}(\Upsilon,\Theta)^{(a)}$ are the scaling functions associated with
the model $(a)$.  The nonuniversal rescaling coefficients $c_\theta$
and $c_y$ do not depend on the particular observable considered, thus
they can be obtained by straightforward matching of the data of $R$
and $U$.  Given that such rescaling coefficients are unique, i.e. they
must be the same for any observable, one may first obtain $c_y$ by
matching equilibrium data at $t=0$, and then determine $c_\theta$ by
matching the out-of-equilibrium data at fixed $\Upsilon$.

We finally mention that analogous scaling ansatzes have been already
put forward, and verified numerically, at quantum
transitions~\cite{PRV-18,RV-21,TV-22}. At quantum transitions, the
equilibrium universality class of the quantum critical behavior can be
inferred using the quantum-to-classical
mapping~\cite{Sachdev-book,RV-21}, while the dynamic exponent $z$ is
associated with the unitary quantum dynamics.

\subsection{The relaxational flow of the energy density}
\label{relscaene}

The out-of-equilibrium FSS behavior outlined in Sec.~\ref{outsca}, for
the RG invariant quantities $R$ and $U$ and the two-point function
$G({\bm x})$, are obtained by a natural extension of their equilibrium
scaling behaviors, adding a time dependence through the time scaling
variable $\Theta$. However, it is not clear whether, and how, this
picture can be extended to the time dependence of quantities
whose equilibrium behavior is affected by analytical backgrounds, such
as the energy density.  As discussed in Sec.~\ref{equfss}, the scaling
term of the equilibrium energy density is generally suppressed with
respect to the analytical background,
cf. Eq.~(\ref{leadINGFSSene}). Moreover, when the specific-heat
exponent $\alpha$ is negative, the subtraction of the critical value,
cf. Eq.~(\ref{diffe}), is not sufficient to make the scaling term
dominant.

Concerning the out-of-equilibrium behavior along the critical
relaxational flow, the question is whether the analytical background
shares the same time scale $\tau\sim L^z$ of the critical modes, or
their time scale $\tau_I$ is significantly shorter, as may be
suggested by the the fact they are expected to arise from short-ranged
fluctuations. If the ratio $\tau_I/\tau$ vanishes in the large-$L$
limit, then the out-of-equilibrium energy density may develop an
out-of-equilibrium scaling behavior characterized by the same power
law of the scaling term of the equilibrium energy density,
cf. Eq.~(\ref{leadINGFSSene}).

To investigate the out-of-equilibrium behavior of the energy density
under critical relaxational flow, we consider the subtracted
post-quench energy density
\begin{eqnarray}
  E_s(t,r,L) \equiv E(t,r,L) - E_c,
  \label{diffet}
\end{eqnarray}
where $E_c$ is the critical value of the energy density in the
thermodynamic limit. Asymptotically, due to the thermalization at the
critical point (guaranteed by the relaxational flow for sufficiently
large times at finite volume), it is suppressed as
$E_s(t\to\infty)\sim L^{-(d-y_r)}$, according to
Eq.~(\ref{leadINGFSSenesub}).

When the specific-heat exponent $\alpha$ is positive, the
$O(L^{y_r-d})$ scaling term is already dominant at the equilibrium
starting point ($t=0$). Therefore, a natural hypothesis is that the
scaling behavior persists along the relaxational flow, i.e.
\begin{eqnarray}
E_s(t,r,L) \approx L^{-(d-y_r)} {\cal E}_s(\Upsilon,\Theta),
  \label{leadINGFSSenesubt}
\end{eqnarray}
arising smoothly from the equilibrium scaling term in
Eq.~(\ref{leadINGFSSenesub}), i.e.  ${\cal E}_s(\Upsilon,\Theta\to 0)
\to {\cal E}_e(\Upsilon)$, like the RG invariant quantities $R$ and
$U$, cf. Eq.~(\ref{gxscadyn}).  The situation appears much less clear
in case the specific-heat exponent $\alpha$ is negative, when at the
starting equilibrium condition the dominant contribution comes from
the regular term. Therefore, we may have that there is no longer a
leading out-of-equilibrium scaling such as
Eq.~(\ref{leadINGFSSenesubt}), or it is somehow recovered for dynamic
reasons.

If the time scales of the thermalization of the different equilibrium
contributions differ substantially, then the time evolution along the
critical relaxational flow may disentangle the scaling contribution
from the terms arising from the mixing with the identity.  In other
words, if the time scale $\tau_I$ of the mixings with the identity is
significantly shorter than $\tau\sim L^z$, so that $\tau_I/\tau\to 0$
in the large-$L$ limit, then the modes contributing to the regular
part should thermalize much earlier than the critical modes. This may
allow the subtracted energy density $E_s(t,r,L)$ to develop an
out-of-equilibrium scaling along the critical relaxational flow even
in the case of negative specific-heat exponent.  More precisely, its
post-quench out-of-equilibrium scaling behavior may be still written
in terms of the scaling variables $\Upsilon$ and $\Theta$, as in
Eq.~(\ref{leadINGFSSenesubt}).

We remark that the fast thermalization of the regular part of the
energy is a reasonable hypothesis, due to the fact that the mixing
with the identity operator is generally expected to arise from
short-ranged modes. However, even this quite simple scenario may not
be so straightforward.  In particular, a potential singularity of the
scaling function ${\cal E}_s(\Upsilon,\Theta)$ can be expected in the
$\Theta\to 0$ limit.  Therefore it is not smoothly related to the
equilibrium scaling function ${\cal E}_e(\Upsilon)$ entering
Eq.~(\ref{leadINGFSSenesub}), due to the fact that it does not provide
the leading behavior at equilibrium when $\alpha<0$.

We finally note that if the energy density shows an out-of-equilibrium
scaling behavior, its scaling function ${\cal E}_s$ is expected to be
a universal function, apart from a multiplicative factor, and from the
nonuniversal normalizations of the scaling variables $\Theta$ and
$\Upsilon$, which can be determined by the matching procedure defined
by Eq.~(\ref{rescaling}).

In the following section we numerically check these out-of-equilibrium
scaling behaviors within the 3D $N$-vector model with $N=3$ and $N=4$,
which provide examples of equilibrium critical behaviors where the
subtracted energy density $E_s$ is dominated by the analytical
background, due to the fact that their specific-heat critical exponent
$\alpha$ is negative.  As we shall see, the numerical results show the
emergence of an out-of-equilibrium scaling behavior even in these
cases, such as the one in Eq.~(\ref{leadINGFSSenesubt}), showing also
a singular diverging behavior of the scaling function ${\cal E}_s$ in
the $\Theta\to 0$ limit.

\section{Numerical results}
\label{numres}

\subsection{MC simulations}
\label{mcsim}

To investigate the out-of-equilibrium behavior of spin systems arising
from the quench protocol outlined in Sec.~(\ref{protflow}), we present
FSS analyses of MC simulations for the $N=3$ and $N=4$ vector models
defined by the Hamiltonian Eq.~(\ref{onmode}).

We generate an ensemble of equilibrium configurations for
$\beta<\beta_c$, obtained by standard MC simulations using overrelaxed
algorithms, see e.g. Ref.~\cite{Hasenbusch-22}, constructed by mixing
heat-bath and microcanonical upgradings (usually mixed with a
probability 1:9).  With intervals suffciently large to achieve a
substantial decorrelation of the equilibrium configurations, we start
MC simulations at $\beta_c$ to generate a corresponding ensemble of
critical relaxational flows, where we use the heat-bath algorithm only
(we use a checkerboard upgrading scheme, where we separately upgrade
the odd and even spin sites).  The resulting evolution corresponds to
a purely relaxational dynamics (model A). The time unit $t=1$
corresponds to a complete checkerboard passage, where all sites are
visited. The fixed-time observables that we consider are measured
along the relaxational flow at fixed time, and averaged over the
ensemble of relaxational flows.

\subsection{Out-of-equilibrium FSS of the RG invariant observables}
\label{resrginv}

\begin{figure}[tbp]
  \includegraphics*[scale=\graphicscale]{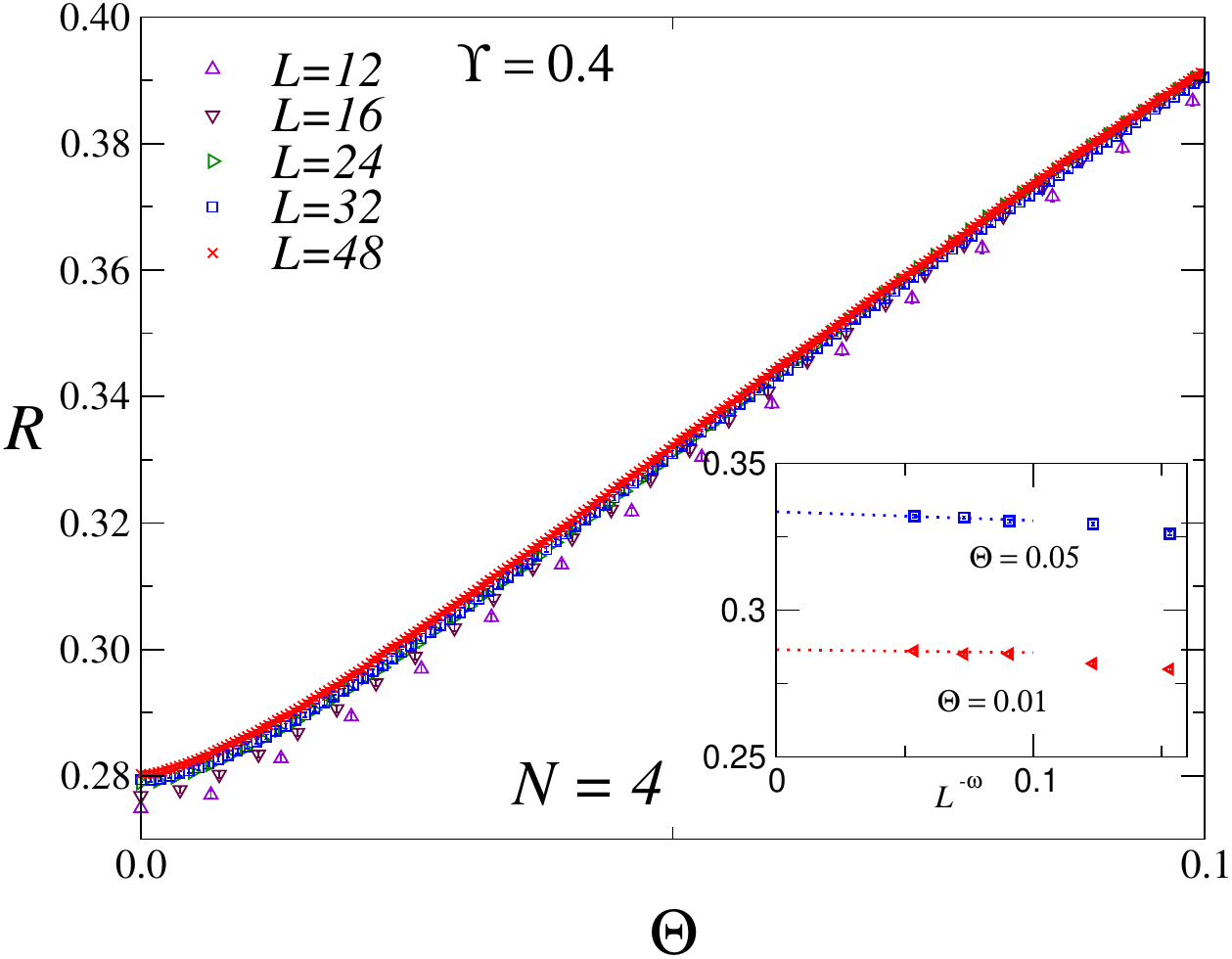}
  \includegraphics*[scale=\graphicscale]{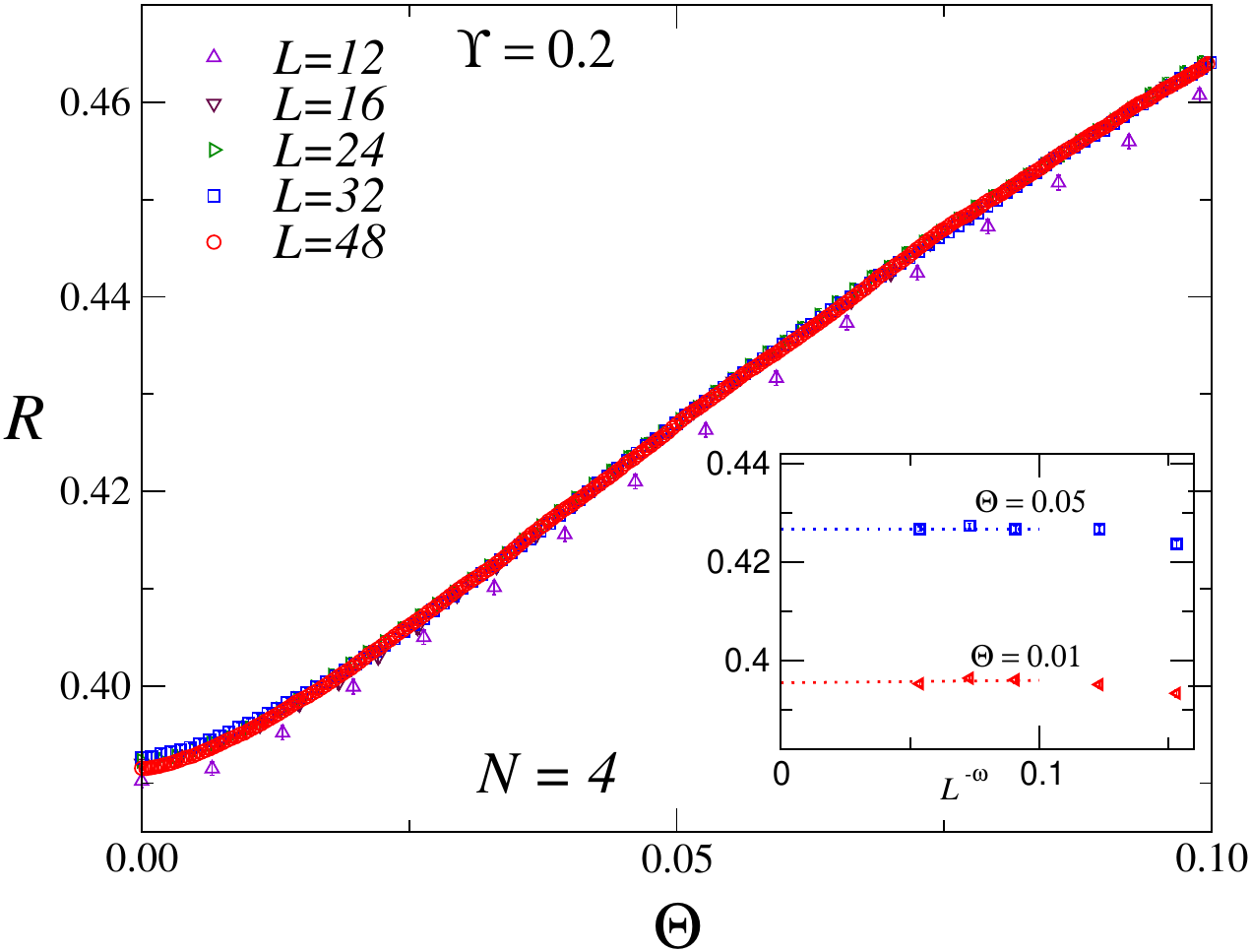}
  \caption{The ratio $R\equiv \xi/L$ along the critical relaxational
    flow versus $\Theta\equiv t/L^z$ for $N=4$, at fixed
    $\Upsilon=0.2$ (bottom) and $\Upsilon=0.4$ (top).  The insets show
    the large-$L$ convergence for some fixed values of $\Theta$,
    plotting the corresponding data versus $L^{-\omega}$, which is the
    expected behavior of the scaling corrections.  }
\label{rxio4}
\end{figure}

\begin{figure}[tbp]
  \includegraphics*[scale=\graphicscale]{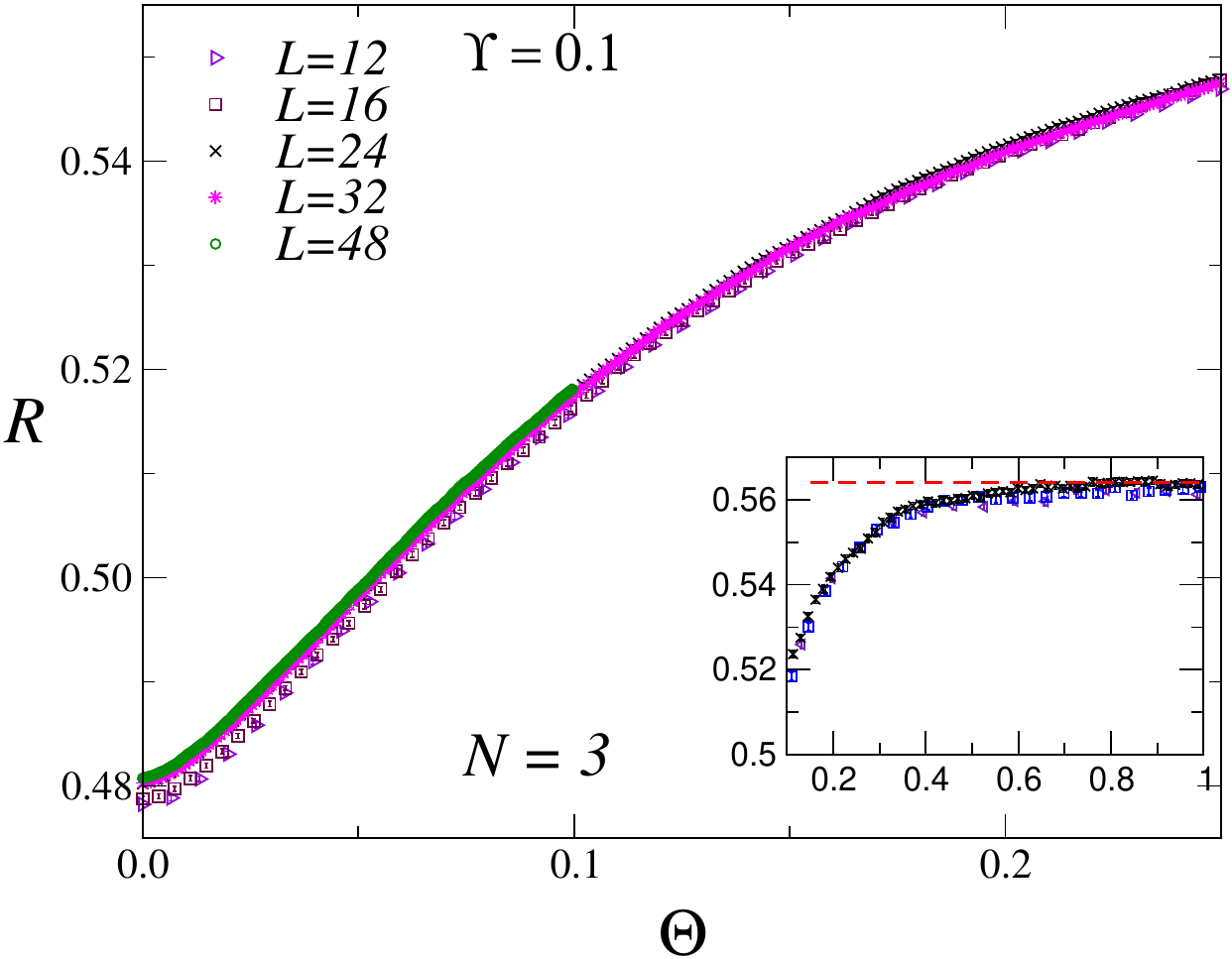}
  \includegraphics*[scale=\graphicscale]{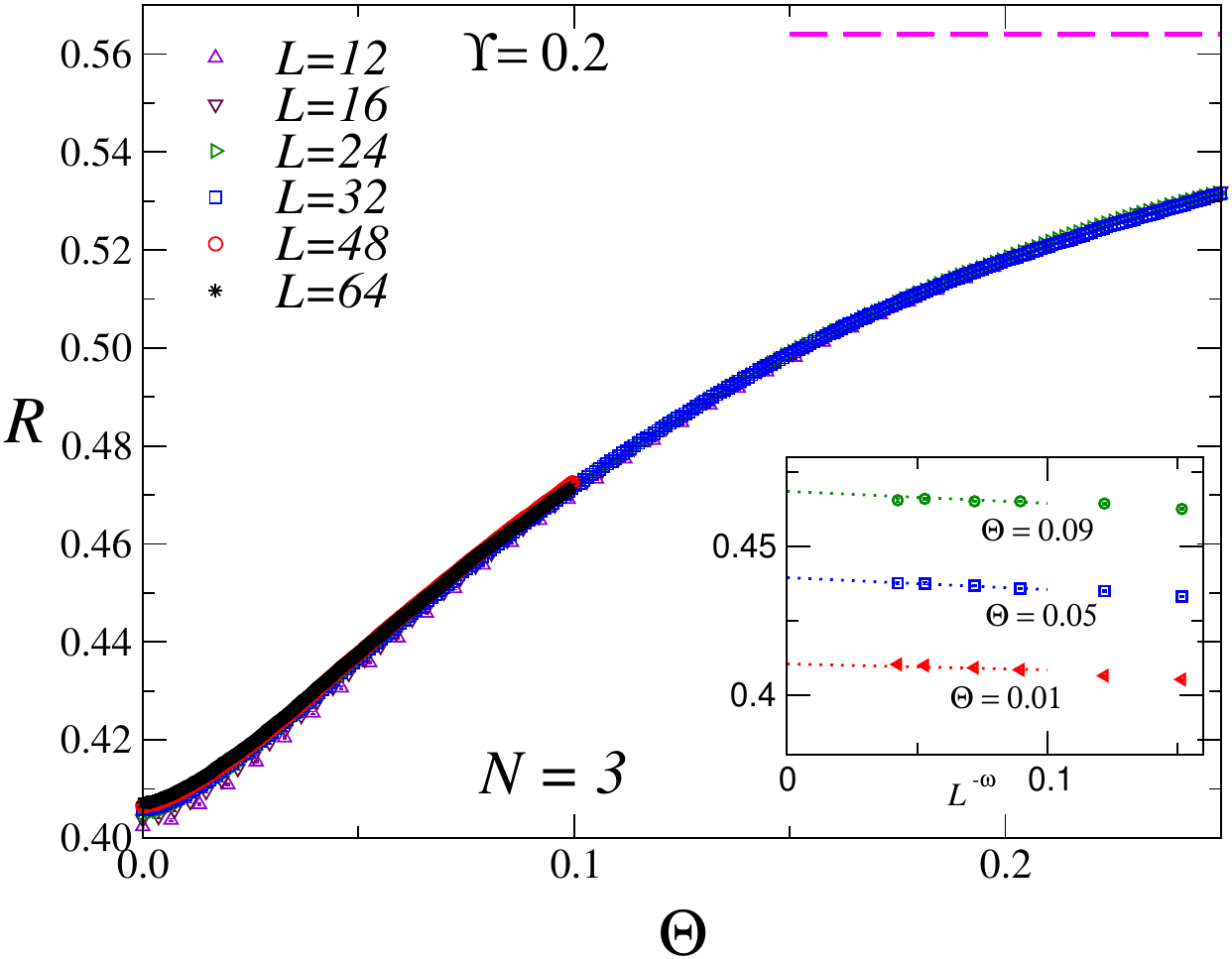}
  \caption{ The ratio $R\equiv \xi/L$ along the critical relaxational
    flow versus $\Theta\equiv t/L^z$ for $N=3$, at fixed
    $\Upsilon=0.2$ (bottom) and $\Upsilon=0.1$ (top).  The inset of
    the bottom figure shows the large-$L$ convergence for some fixed
    values of $\Theta$, versus their expected leading $O(L^{-\omega})$
    suppression. The inset of the top figure shows the data up to a
    relatively large value $\Theta=1$, for lattice sizes up
    to $L=24$, demonstrating the large time convergence to the
    corresponding equilibrium value at the critical point, which, in
    turn, converges to the critical value~\cite{Hasenbusch-20-o3}
    $R^*=0.56404(2)$ in the large-$L$ limit (indicated by the dashed
    line)}
\label{rxio3}
\end{figure}

We first show results for the RG invariant quantity $R\equiv \xi/L$
along the critical relaxational flow, arising from the protocol
outlined in Sec.~\ref{protflow}. Analogous results are obtained for
the Binder parameter $U$, therefore we do not show them because they
do not add further relevant information.

In Figs.~\ref{rxio4} and \ref{rxio3} we report results for the ratio
$R\equiv \xi/L$ at fixed values of $\Upsilon \equiv r L^{y_r}$ (where
$y_r=1/\nu$) respectively for the $N=4$ and $N=3$ vector models. Their
plots versus the time scaling variable $\Theta=t/L^z$ fully confirm
the out-of-equilibrium FSS behavior reported in
Eq.~(\ref{gxscadyn}). Indeed, the data appear to approach an
asymptotic scaling curve, in the large-$L$ and large-$t$ limit keeping
$\Upsilon$ and $\Theta$ constant. The scaling corrections appear quite
small; their suppression is consistent with the $O(L^{-\omega})$
behavior already present at equilibrium.

The asymptotic scaling curve starts from the equilibrium value ${\cal
  R}(\Upsilon,\Theta=0) = {\cal R}_e(\Upsilon)$ and appears to
approach the asymptotic large-time behavior ${\cal
  R}(\Upsilon,\Theta\to\infty) = {\cal R}_e(0)$, whose value is known
with great accuracy, i.e.~\cite{Hasenbusch-20-o3} ${\cal
  R}_e(0)=0.56404(2)$ for $N=3$ and~\cite{Hasenbusch-22} ${\cal
  R}_e(0)=0.547296(26)$ for $N=4$.  Some results for the large time
approach are shown in the inset of the top Fig.~\ref{rxio3} for the
lattice size up to $L=24$, showing clearly the convergence to the
corresponding equilibrium value. Analogous results (not reported) are
obtained for the Binder parameter $U$, cf. Eq.~(\ref{defU}).

\subsection{Out-of-equilibrium scaling of the energy density}
\label{resenedens}

\begin{figure}[tbp]
  \includegraphics*[scale=\graphicscale]{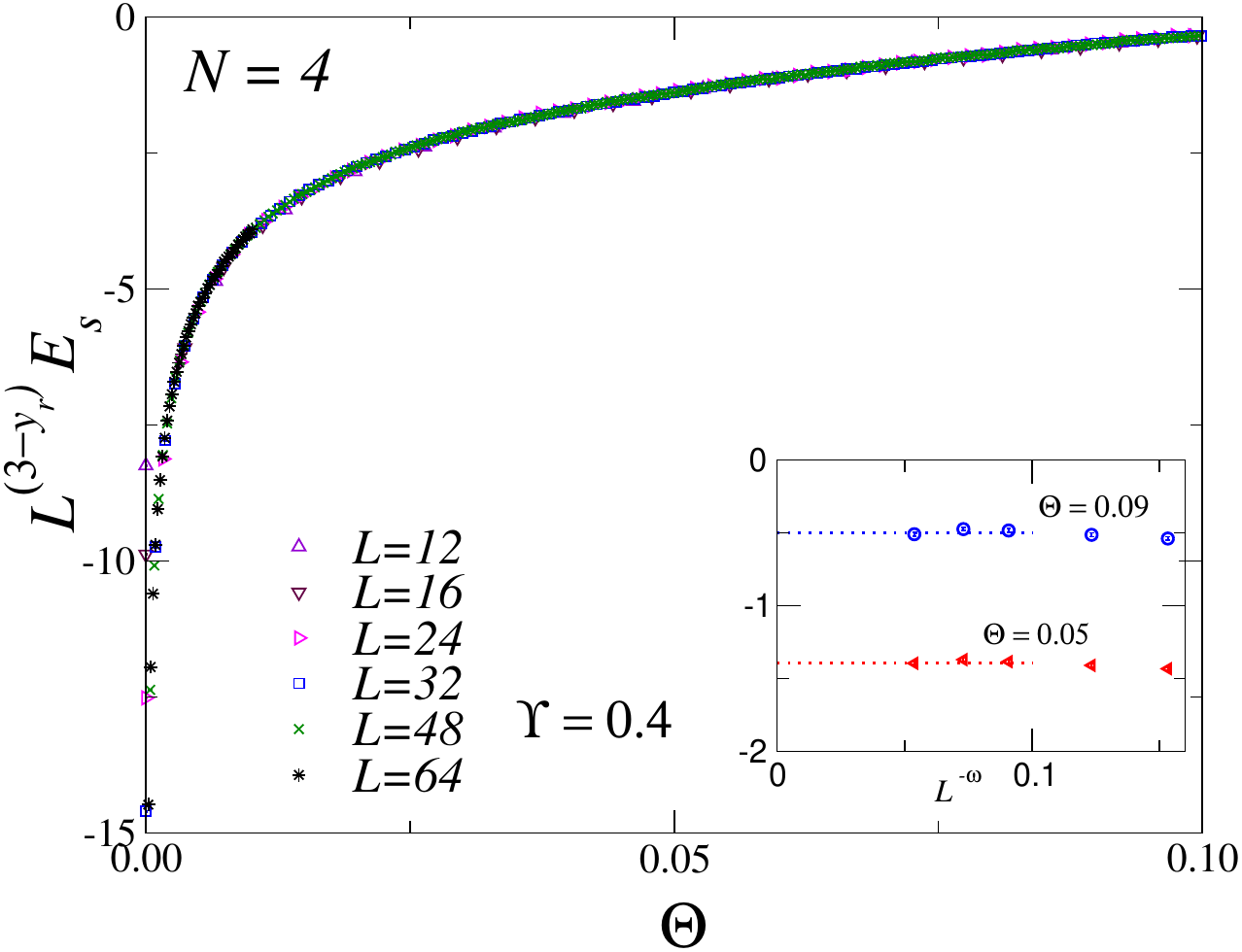}
  \includegraphics*[scale=\graphicscale]{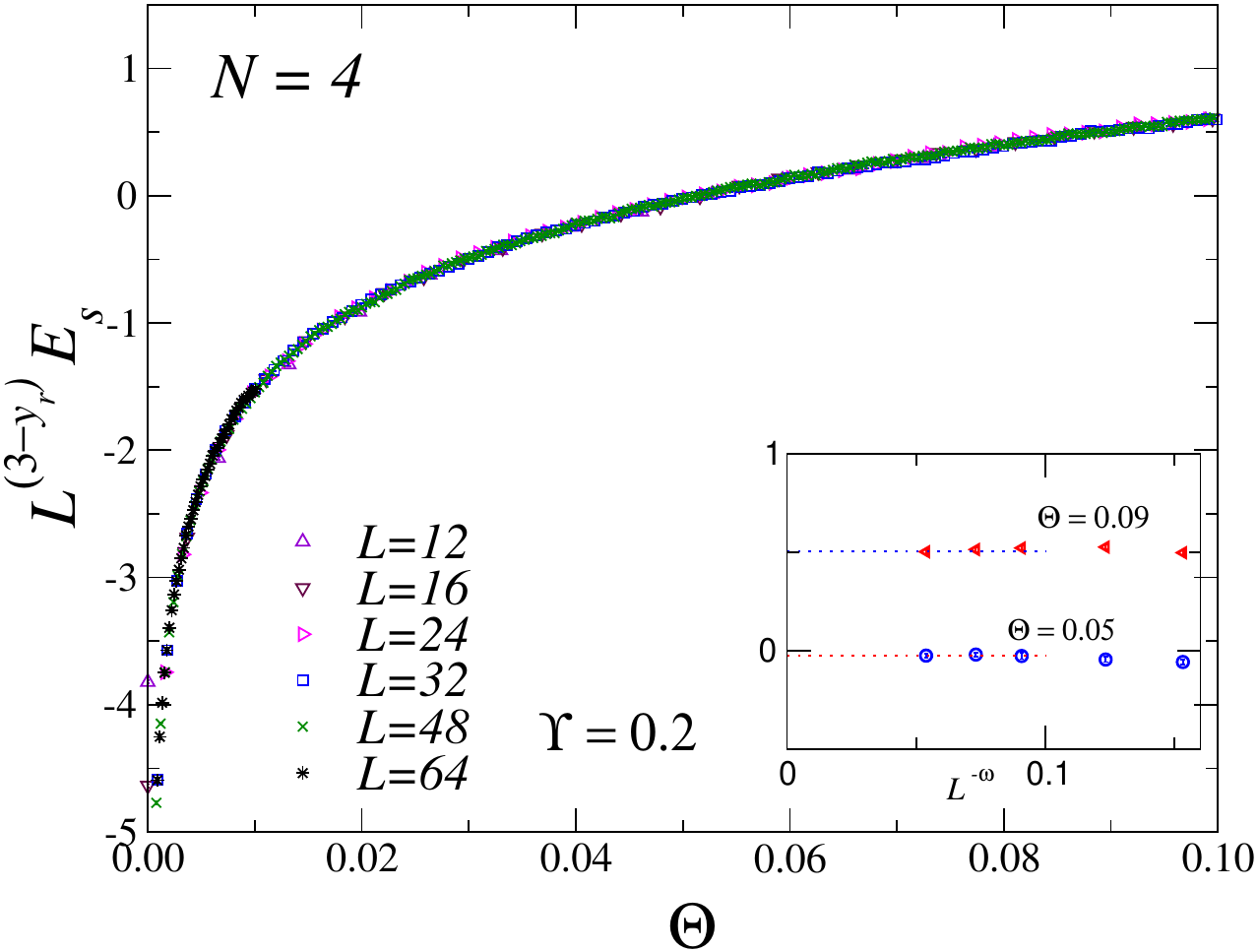}
  \caption{Out-of-equilibrium FSS of the subtracted energy density
    $E_s$ along the critical relaxational flow for $N=4$, at fixed
    $\Upsilon=0.2$ (bottom) and $\Upsilon=0.4$ (top).  The insets show
    the large-$L$ convergence for some fixed values of $\Theta$,
    indicating that the scaling corrections are small, and consistent
    with $O(L^{-\omega})$. Note that the data at $t=0$ do not scale,
    see Fig.~\ref{t0enedata}.}
\label{eneo4}
\end{figure}

\begin{figure}[tbp]
  \includegraphics*[scale=\graphicscale]{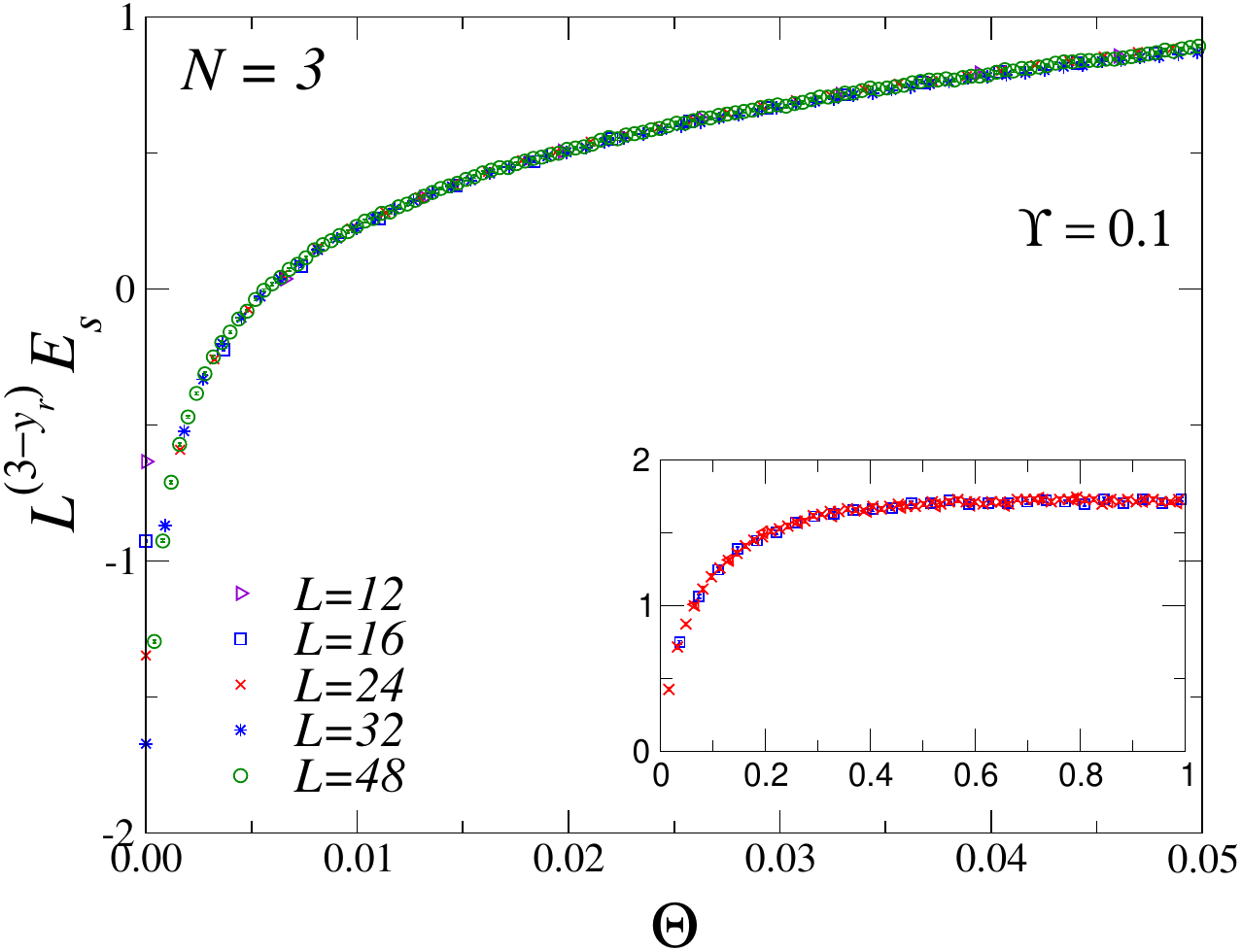}
  \includegraphics*[scale=\graphicscale]{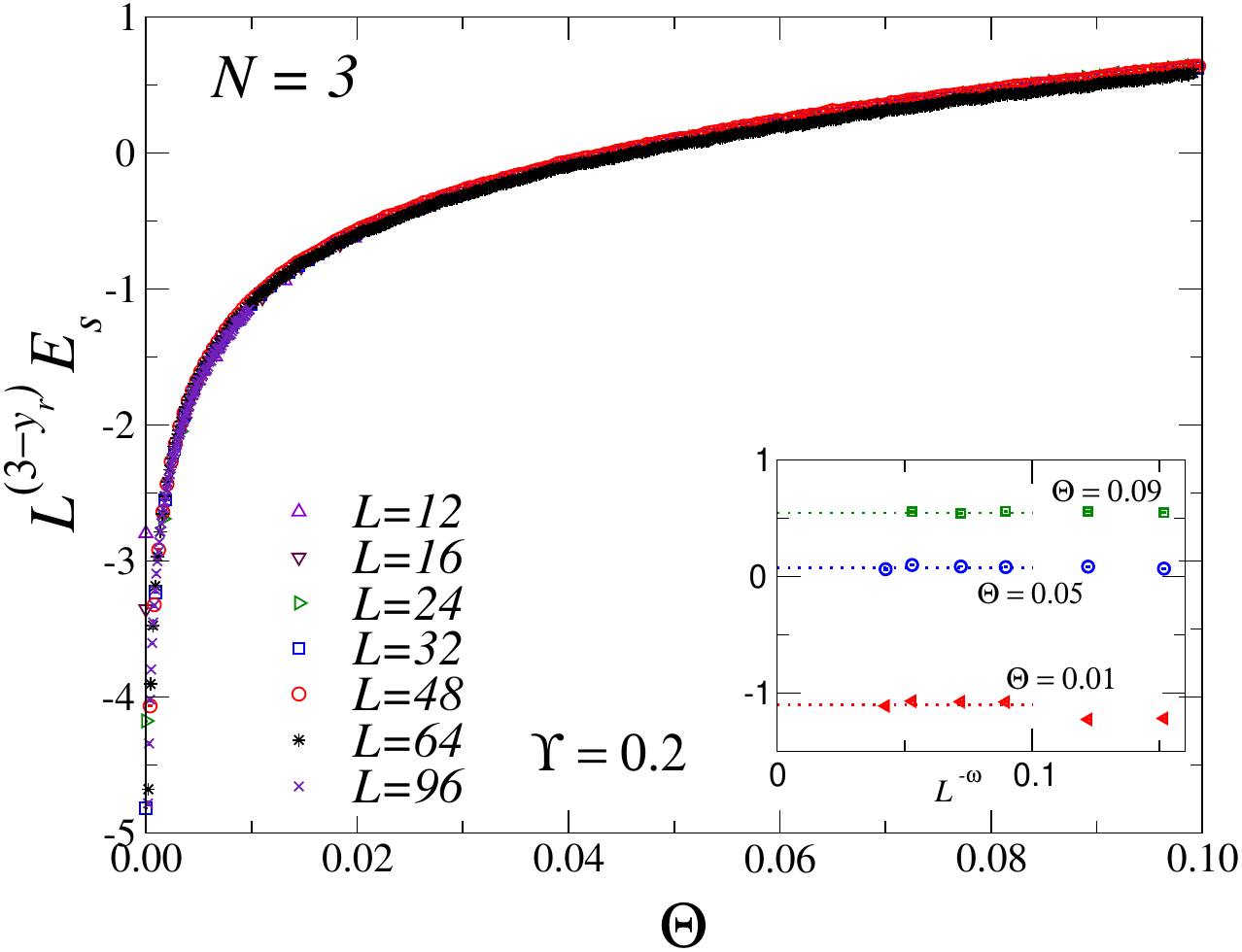}
  \caption{Out-of-equilibrium FSS of the $N=3$ subtracted energy
    density $E_s$ along the critical relaxational flow, at fixed
    $\Upsilon=0.2$ (bottom) and $\Upsilon=0.1$ (top).  The inset of
    the bottom figure shows the large-$L$ convergence for some fixed
    values of $\Theta$.  The inset of the top figure shows the data up
    to a relatively large value of $\Theta$, i.e. $\Theta=1$, at fixed
    $\Upsilon=0.1$ and for lattice sizes up to $L=24$, demonstrating
    the large time convergence to the corresponding equilibrium value
    at the critical point.  Note that the data at $t=0$ do not scale,
    see Fig.~\ref{t0enedata}.}
\label{eneo3}
\end{figure}

To investigate the out-of-equilibrium scaling behavior of the
subtracted energy density $E_s$, cf. Eq.~(\ref{diffet}), we first need
to obtain accurate estimates of the energy density $E_c$ at the
critical point in the infinite volume limit.  By standard MC
simulations at the critical point, up to $L=128$, we obtain the quite
accurate estimates $E_c=0.989498(7)$ for $N=3$, and $E_c=0.991708(3)$
for $N=4$. They are obtained by fitting the data to the expected
large-$L$ behavior: $E(\beta_c,L) = E_c + b L^{-(3-y_r)}$, and also
$E(\beta_c,L) = E_c + b L^{-(3-y_r)}(1+cL^{-\omega})$ to include the
leading scaling correction term.

We now analyze the behavior of the subtracted energy density along the
critical relaxational flow arising from the protocol outlined in
Sec.~\ref{protflow}. In Figs.~\ref{eneo4} and \ref{eneo3} we show the
results for the $N=4$ and $N=3$ vector models, respectively, for some
fixed values of $\Upsilon$, up to quite large lattice sizes.  They
clearly support the asymptotic FSS behavior proposed in
Eq.~(\ref{leadINGFSSenesubt}), in the large-$L$ and large-$t$ limit
keeping $\Upsilon$ and $\Theta$ constant.  Indeed, the plots of the
the product $L^{d-y_r}E_s(t,r,L)$ versus $\Theta$, at fixed
$\Upsilon$, clearly approach a scaling curve ${\cal
  E}_s(\Upsilon,\Theta)$ when increasing $L$.  Again the scaling
corrections to the asymptotic behavior appear small and consistent
with an $O(L^{-\omega})$ approach.

\begin{figure}[tbp]
  \includegraphics*[scale=\graphicscale]{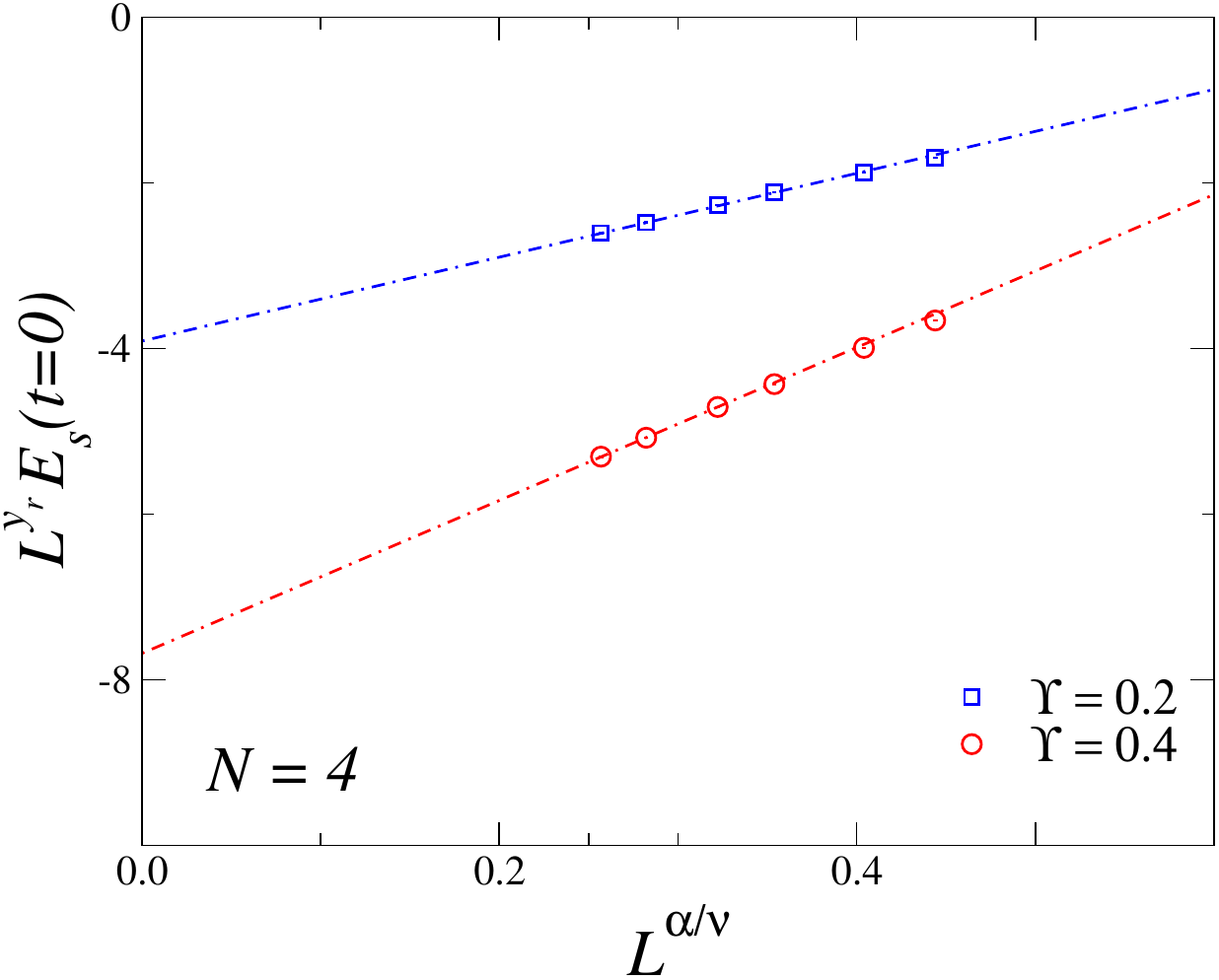}
  \caption{The subtracted energy $E_s$ at $t=0$ for $N=4$ and some
    values of $\Upsilon$. The error of the data are practically
    invisible in the plot.  To evidentiate the expected equilibrium
    behavior reported in Eqs.~(\ref{leadINGFSSenesub}) and
    (\ref{regeexp}), we plot $L^{y_r} E_s(t=0)$ versus
    $L^{\alpha/\nu}$ with $\alpha/\nu=2/\nu - d \approx -0.19$, which
    is the relative power law of the scaling term with respect to the
    background contribution at equilibrium. The data show the behavior
    reported in Eq.~(\ref{est0}).  The lines show linear fits of the
    data for the larger lattices, which have acceptable $\chi^2$.  We
    stress that this scaling behavior does not match that observed at
    fixed $\Theta>0$. Analogous power-law behaviors are observed for
    $N=3$.}
\label{t0enedata}
\end{figure}

Of course, the data at $t=0$, corresponding to the initial equilibrium
conditions, do not appear to scale in the plots shown in
Fig.~\ref{eneo4} and \ref{eneo3}, because they are expected to behave
as $E_{se}\sim L^{-y_r}$, cf. Eqs.~(\ref{leadINGFSSenesub}) and
({\ref{regeexp}), due to the fact that the leading contribution comes
  from the regular term $\Delta E_{\rm reg} \sim r$, and $r\sim
  L^{-y_r}$ when keeping $\Upsilon$ fixed. This is clearly shown by
  the plots in Fig.~\ref{t0enedata} of the $t=0$ data for $N=4$ and
  fixed $\Upsilon$, which confirm that
\begin{equation}
  E_s(t=0,r,L) = b_1 \Upsilon L^{-y_r} \left[ 1 + O(L^{2y_r-d}) \right],
\label{est0}
\end{equation}
where $2y_r-d=\alpha/\nu<0$.  Therefore, the quantity plotted in
Figs.~\ref{eneo4} and \ref{eneo3} diverges as $L^{d-y_r} E_s \sim
L^{-\alpha/\nu}$ at $t=0$.

Nevertheless, we note that the out-of-equilibrium FSS is nicely
observed at very small values of $\Theta$ already.  It is important to
observe that a closer analysis of the data in Figs.~\ref{eneo4} and
\ref{eneo3} at small values of $\Theta$ suggest a singular diverging
behavior of the scaling functions.  In Fig.~\ref{logene} we illustrate
the behavior at small $\Theta$ by log-log plots.  As already
anticipated in Sec.~\ref{relscaene}, the limit $\Theta\to 0$ appears
to be singular, essentially because we do not expect a smooth approach
to the equilibrium behavior, where its scaling behavior gets
suppressed with respect to the leading regular term, cf.
Eq.~(\ref{leadINGFSSenesub}).  The data seem to confirm a singular
diverging behavior for $\Theta\to 0$, likely as ${\cal
  E}_s(\Upsilon,\Theta\to 0)\sim \Theta^{-\kappa}$.  The value of
$\kappa$ may be predicted by matching the lattice-size power law in
the $\Theta\to 0$ limit with the leading $O(L^{-y_r})$ term at $t=0$,
obtaining
\begin{equation}
  \kappa = - {\alpha\over z\nu}>0.
  \label{kappaval}
\end{equation}  
As shown in Fig.~\ref{logene}, this power-law behavior is confirmed by
the fitting the data of $L^{d-y_r} E_s$ at small $\Theta$,
i.e. $\Theta\lesssim 0.002$, to the asymptoptic $\Theta\to 0$ behavior
\begin{equation}
{\cal E}_s(\Theta,\Upsilon) \approx
A(\Upsilon) \Theta^{-\kappa} + B(\Upsilon).
\label{essmth}
\end{equation}
For example, for the data at $\Upsilon=0.2$ shown in
Fig.~\ref{logene}, we obtain $A\approx -3.1$ and $B\approx 5.1$ for
$N=4$, and $A\approx -5.4$ and $B\approx 7.2$ for $N=3$.  

On the other hand, the approach to the $\Theta\to\infty$ limit appears
smooth, as shown by the inset of the top Fig.~\ref{eneo3} where we
show extended data for $N=3$ up to larger times, i.e. $\Theta=1$.
This demonstrates the regular convergence to the asymptotic critical
equilibrium scaling behavior, where the subtracted energy density is
expected to scale as $E_s(t\to\infty)\sim L^{-(d-y_r)}$ due the
subtraction of the large-$L$ critical-point value $E_c$.

\begin{figure}[tbp]
  \includegraphics*[scale=\graphicscale]{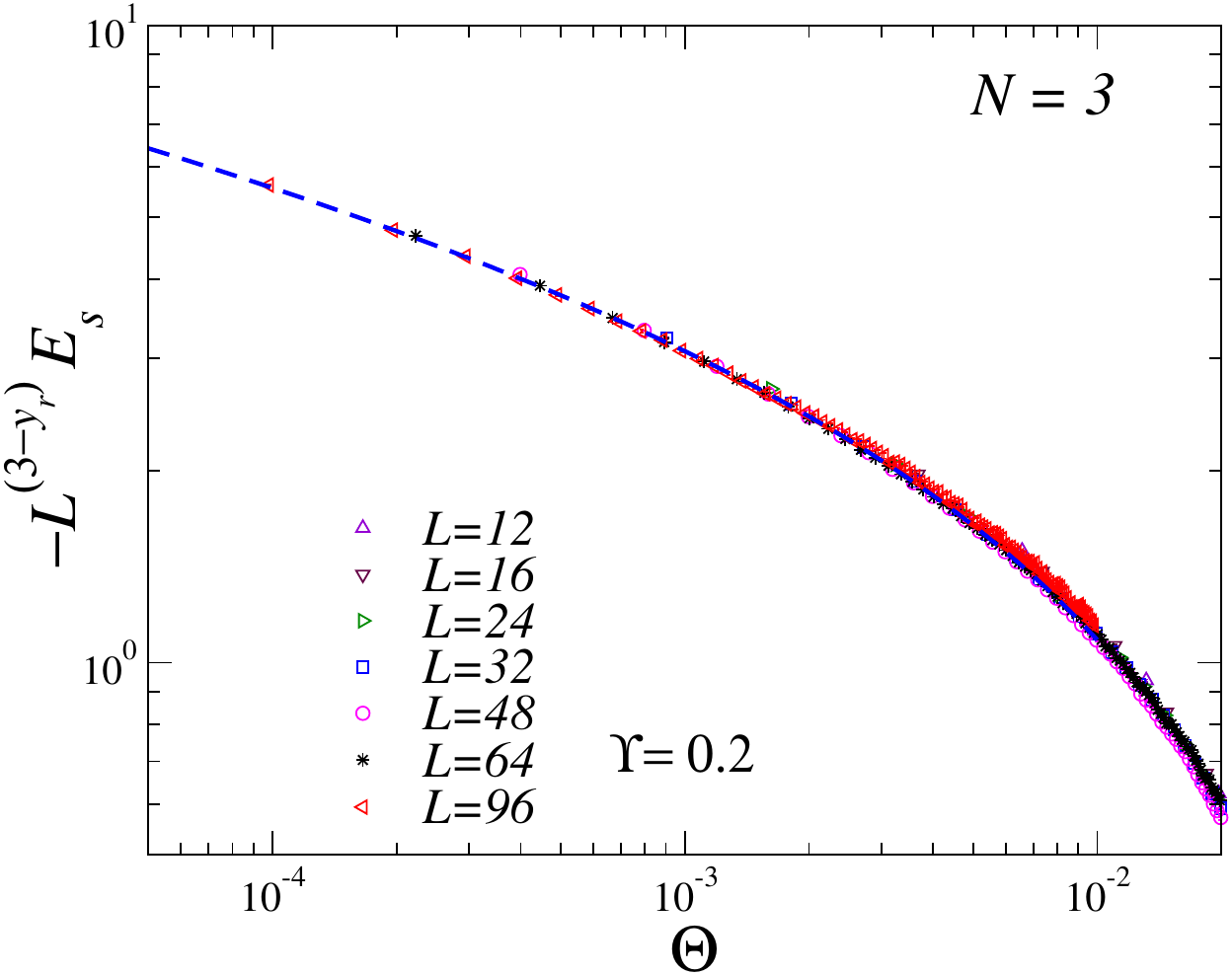}
    \includegraphics*[scale=\graphicscale]{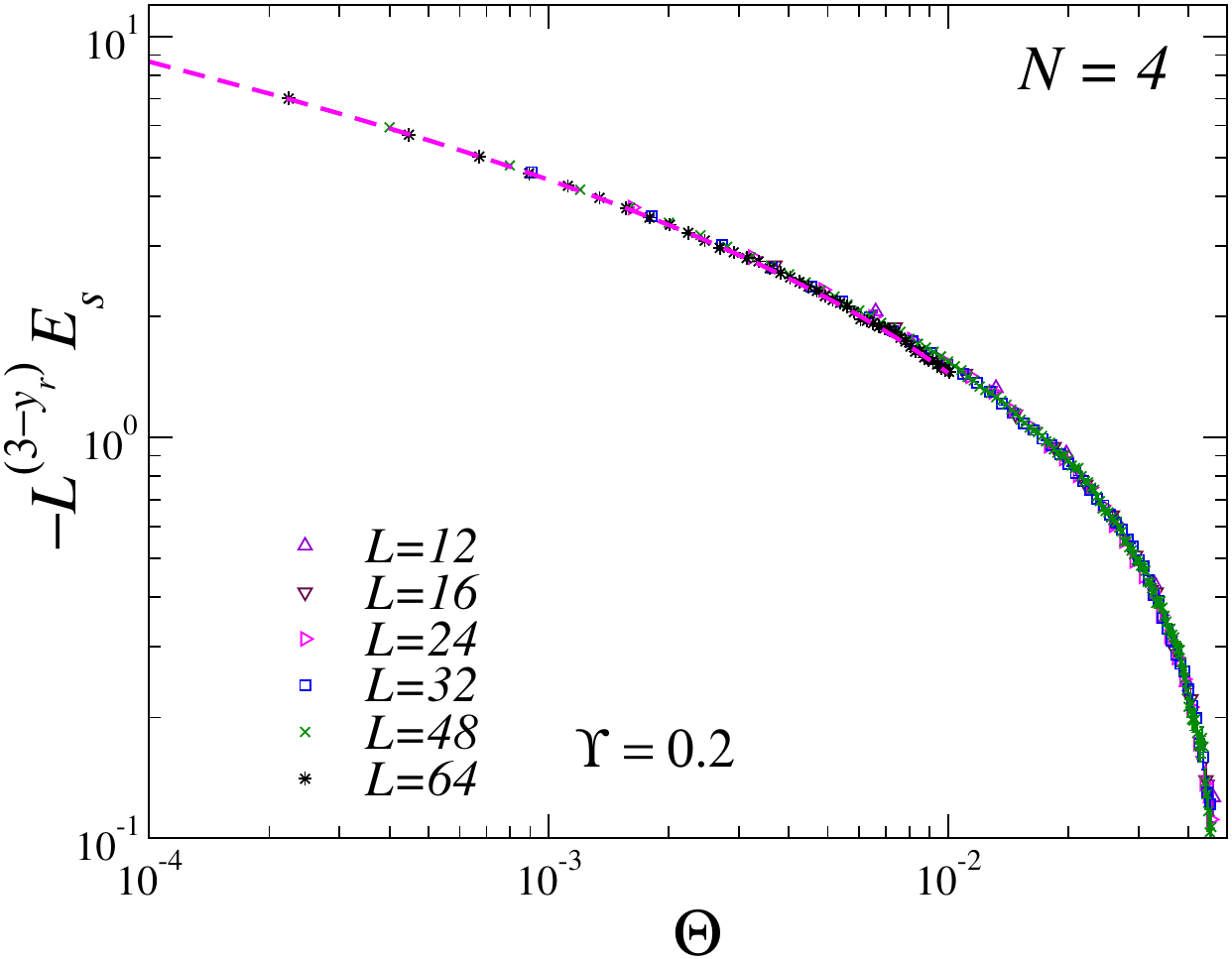}
  \caption{We show a log-log plot of data of the subtracted energy
    density to make evident the behavior at small $\Theta$, for $N=3$
    (top) and $N=4$ (bottom), at $\Upsilon=0.2$. The dashed lines show
    linear fits of the data of $L^{d-y_r} E_s$ at small $\Theta$,
    i.e. $\Theta\lesssim 0.002$, to the asymptoptic $\Theta\to 0$
    behavior $A\Theta^{-\kappa} + B$ with $\kappa=-\alpha/(z\nu)$.
    Analogous results are obtained for other values of $Y$, such as
    $Y=0.4$ for $N=4$ and $Y=0.1$ for $N=3$.}
\label{logene}
\end{figure}

\section{Conclusions}
\label{conclu}

We have addressed the out-of-equilibrium critical behavior of
statistical systems subject to a quenching protocol of the
temperature, giving rise to a critical relaxational flow at the
critical point, starting from equilibrium conditions outside
criticality, as outlined in Sec.~\ref{protflow}.  We focus on the
out-of-equilibrium scaling behaviors emerging in soft temperature
quenches, i.e. when the temperature variation associated with the
quench is sufficiently small to maintain the system close to
criticality along the whole post-quench out-of-equilibrium
dynamics~\cite{RV-21}.

The out-of-equilibrium scaling behaviors associated with soft-quench
protocols can be described by extensions of the equilibrium scaling
predicted by the RG theory, obtained by adding a further dependence of
the scaling functions on the time after quench, through an appropriate
time scaling variable $\Theta\equiv t/L^z$, where $z$ is the universal
dynamic exponent associated with the type of dynamics considered.  As
discussed in the paper, this extension appears quite natural for
observables that have already a well-defined leading equilibrium FSS
behavior, such as the two-point function of the order-parameter
field. In this case one also expects that both equilibrium limits
$\Theta\to 0$ and $\Theta\to \infty$ of their out-of-equilibrium FSS
are regular.

The problem gets more complicated in the case of observables whose
equilibrium behavior at the transition is dominated by a regular
background term~\cite{PV-02}, or mixing with the identity operator, so
that the singular scaling behavior is only subleading at criticality,
thus hidden by the dominant regular contributions. We argue that, if
the time scales of the thermalization of the different equilibrium
contributions differ substantially, then the time evolution along the
critical relaxational flow may somehow disentangle the scaling
contribution from the terms arising from the mixing with the identity,
due to the fact that the modes contributing to the regular part should
thermalize much earlier than the critical modes. This may allow the
subtracted energy density, cf. Eq.~(\ref{diffet}), to develop an
out-of-equilibrium scaling along the critical relaxational flow even
in the case of negative specific-heat exponent, i.e.  when the
equilibrium scaling behavior is still dominated by the regular
background. However, the equilibrium $\Theta\to 0$ limit of the
corresponding out-of-equilibrium scaling functions may develop a
singular (diverging) behavior.

We have numerically investigated this out-of-equilibrium scenario
within the 3D lattice $N$-vector models for $N=3$ and $N=4$, which
provide examples of critical behaviors with negative specific-heat
exponents.  We analyze the out-of-equilibrium behavior along critical
relaxational flows arising from instantaneous quench of the
temperature $T$ at a time $t=0$, from $T>T_c$ to the critical point
$T_c$, starting from equilibrium conditions. Our numerical analyses
within dynamic FSS frameworks show that the energy density also
develops an out-of-equilibrium FSS, after subtraction of its
asymptotic critical value at $T_c$. Moreover, the numerical results
hint at a (likely power-law) diverging $\Theta\to 0$ behavior of the
out-of-equilibrium scaling functions of the subtracted energy density,
unlike other observables closely related to the critical modes such as
the two-point function of the order-parameter field.

It would be important to further confirm this scenario by considering
other models within the same universality classes and type of
dynamics, to check the universality of the out-of-equilibrium FSS of
the energy density during the critical relaxational flow. Moreover,
results for other systems within other universality classes may also
turn out to be very useful to clarify the nature of the $\Theta\to 0$
singularities of the out-of-equilibrium FSS functions of the
subtracted energy density.  Likely, large-$N$ frameworks may provide
an analytical tool to investigate this issue within 3D systems (we
recall that $\alpha=-1$ in the large-$N$ limit of 3D $N$-vector
models).  One may also study issues related to the out-of-equilibrium
behavior of the energy density within other dynamic protocols, such as
protocols entailing slow changes of the temperature, analogous to
those employed to investigate the so-called out-of-equilibrium
Kibble-Zurek problem~\cite{Kibble-80,Zurek-96,CEGS-12}.

We mention that analogous studies may turn out to be interesting also
at quantum transitions, in the zero-temperature limit of quantum
many-body systems~\cite{Sachdev-book}, which also show physically
relevant observables dominated by regular background terms, such as
the transverse magnetization (or equivalently the fermionic particle
density) in quantum Ising models~\cite{RV-21}, In this respect, we
recall that quantum systems are subjected to a unitary
energy-conserving dynamics, unlike the relaxational dynamics of
classical systems.

As already mentioned, the critical relaxational flow is somehow
analogous to the so-called gradient flow widely used in lattice QCD to
define scaling observables, and in particular to define a scaling
coupling constant from the energy
density~\cite{Luscher-10,LW-11,MS-15,MSS-15,HN-16}.  The scaling
behavior of the energy density along the gradient flow was conceived
within perturbative approaches~\cite{Luscher-10,LW-11}, and then
verified numerically. Our results on the critical relaxational flow
within the more general context of critical phenomena may help to
clarify some of the basic assumptions of the applications of the
gradient flow in lattice QCD, and in particular the possibility of
extracting a well defined running coupling constant from the energy
density under the gradient flow, beyond perturbation theory.

\acknowledgments

We thank Claudio Bonati for interesting and useful
discussions. H.P. acknowledges financial support from the Cyprus
Research and Innovation Foundation under contract number
EXCELLENCE/0421/0025.

\end{document}